\documentstyle[aps]{revtex}

\begin{document}

\title{On construction of recursion operator and algebra of symmetries for field
and lattice systems \thanks{Supported by KBN research grant no. P03B 113 13}}
\author{Maciej B\l aszak \\
Physics Department, A. Mickiewicz University, \\
Umultowska 85, 61-614 Pozna\'{n} , Poland\\
e-mail: blaszakm@main.amu.edu.pl}
\maketitle

\begin{abstract}
In the paper, developing the idea of V.Sokolov et all. (J.Math.Phys. 40
(1999) 6473) we construct recursion operators and hereditary algebra of
symmetries for many field and lattice systems.
\end{abstract}

\section{ Introduction}

In the investigation of nonlinear integrable field and lattice systems in
(1+1)-dimensions an important role is played by the algebra of symmetries.
For these systems for which additionally exists a so called recursion
operator with vanishing Nijenhuis torsion, the algebra of symmetries takes
the form of the hereditary algebra (centerless Virasoro algebra).
Nevertheless, the construction of the recursion operator of a given
integrable system is not an easy task. Several methods were proposed of
greater  or lower generality, see for example refs. \cite{f},\cite{a},\cite{tu},\cite
{g}. Perhaps the best known is the one based on the implectic-symplectic
factorization of the recursion operator, i.e. on a bi-Hamiltonian property
of the system cosidered. But then the problem is shifted to the construction
of two Poisson structures which is a complicated problem and on a general
level requires advances tools \cite{b1}. Moreover, it fails in the case of non-Hamiltonian reductions.

Recently, amasingly simple and general approach to the construction of
recursion operators was proposed by Sokolov et al. \cite{s}. The method is
based on Lax representation of a given hierarchy of integrable systems and
allows a construction of a recursion operators through really elementary
calculation.

In the paper presented we develop the idea of Sokolov, applying it to
different classes of Lax chains for field and lattice systems. Moreover, we
construct from Lax representation in a systematic way conformal symmetries,
beeing complementary ingredients in the construction of hereditary algebras.

\section{Preliminaries}

Let us consider the following scalar Lax operators (together with their
admissible reductions):

\begin{description}
\item[(i)]  $L=\partial _x^N+u_{N-2}\partial _x^{N-2}+u_{N-3}\partial
_x^{N-3}+...+u_0,$

\item[(ii)]  $L=\partial _x^N+u_{N-1}\partial _x^{N-1}+u_{N-2}\partial
_x^{N-2}+...+u_0+\partial _x^{-1}u_{-1},$

\item[(iii)]  $L=u_N^N\partial _x^N+u_{N-1}\partial _x^{N-1}+u_{N-2}\partial
_x^{N-2}+...+u_0+\partial _x^{-1}u_{-1}+\partial _x^{-2}u_{-2},$
\end{description}

for field systems and

\begin{description}
\item[(iv)]  $L={\cal E}^{N+\alpha }+u_{N+\alpha -1}{\cal E}^{N+\alpha
-1}+...+u_0+u_{-1}{\cal E}^{-1}+...+u_\alpha {\cal E}^\alpha
,\,\,\,\,\,$
\end{description}

for lattice systems, where $-N<\alpha \leq -1,\,\,\,N+\alpha \geq 1.$ \newline
Here and further on we use the following notation for differential and shift
operators 
\[
\partial _xa(x)=a_x+a\partial _x,\,\,\,\,D_xa(x)=a_x, 
\]
\[
{\cal E}a(n)=a(n+1){\cal E},\,\,\,Ea(n)=a(n+1). 
\]
The related Lax equations are 
\begin{eqnarray*}
(i)\,\,\,L_{t_q} &=&\left[ L^{\frac qN}\,_{\geq 0},L\right] , \\
&& \\
(ii)\,\,\,L_{t_q} &=&\left[ L^{\frac qN}\,_{\geq 1},L\right] , \\
&& \\
(iii)\,\,\,L_{t_q} &=&\left[ L^{\frac qN}\,_{\geq 2},L\right] , \,\,\,\,\,\,\,\,\,\,\,\,\,\,\,\,\,\,\,\,q=1,2,...\\
&& \\
(iv)\,\,\,L_{t_q} &=&\left[ L^{\frac q{N+\alpha }}\,_{\geq 0},L\right] ,
\end{eqnarray*}
where $L^{\frac qN}$ is a pseudodifferential series of the form $L^{\frac
qN}=\sum_{-\infty }^qv_i\partial _x^i$ and $L^{\frac qN}\,_{\geq
r}=\sum_r^qv_i\partial _x^i.$ Here $v_i$ are some concrete functions
depending on the coefficients of $L$.

Fixing $N$ (or $N$ and $\alpha $), each case gives an infinite hierarchy of
commuting flows. The case $(i)$ was first considered by Gelfand and Dikii 
\cite{gd}, cases $(ii),(iii)$ by Kupershmidt \cite{k1} and the case $(iv)$
by B\l aszak and Marciniak \cite{b2}, respectively. For arbitrary flow from $%
(i)-(iv)$ we are going to construct a respective algebra of symmetries. What
we really need for this construction are two invariant objects, i.e. the
related conformal (scaling) vector field $\sigma $ and a recursion operator $%
\phi $%
\begin{equation}
{\cal L}_K\sigma +\frac{\partial \sigma }{\partial t}=0,\,\,\,\,\,%
{\cal L}_K\phi =0,  \label{1}
\end{equation}
where $K$ is a vector field of a given flow and ${\cal L}$ means a Lie
derivative.
\newline
{\em Lemma 1}

If 
\[
{\cal L}_\sigma K=\rho K,\,\,\,\,\,\,{\cal L}_\sigma \phi =\alpha \phi
,\,\,\,\,\rho ,\alpha =const,
\]
\begin{equation}
K_n:=\phi ^nK,\,\,\,\,\sigma _n:=\phi ^n\sigma   \label{2}
\end{equation}
and 
\begin{equation}
{\cal L}_{\phi \tau }\phi =\phi {\cal L}_\tau \phi ,  \label{3}
\end{equation}
for an arbitrary vector field $\tau $, then $K_n$ and $\sigma _n$ form a
hereditary algebra 
\begin{equation}
\lbrack K_n,K_m]=0,\,\,\,[\sigma _n,K_m]=(\rho +\alpha
m)K_{n+m},\,\,\,[\sigma _n,\sigma _m]=\alpha (m-n)\sigma _{n+m}.  \label{4}
\end{equation}

The proof the reader can find for example in ref.\cite{b1}. Here we show how
to extract scaling symmetry and a recursion operator directly from Lax
equations.

\section{Scaling properties of Lax equations}

Let us start from scaling symmetries for field systems. Lax equations $%
(i)-(iii) $ are homogenous with respect to the following scaling 
\begin{equation}
\partial _x\rightarrow e^\varepsilon \partial _x,\,\,\,u_{N-i}\rightarrow
e^{i\varepsilon }u_{N-i},\,\,\,\partial _{t_q}\rightarrow e^{q\varepsilon
}\partial _{t_q},\,\,\,L\rightarrow e^{N\varepsilon }L.  \label{5}
\end{equation}
The conformal point transformation takes the form 
\begin{equation}
\varphi _\varepsilon \cdot \left( 
\begin{array}{c}
u_N \\ 
u_{N-1} \\ 
\vdots \\ 
u_0 \\ 
u_{-1} \\ 
u_{-2}
\end{array}
\right) =\left( 
\begin{array}{c}
e^{0\varepsilon }u_N(e^\varepsilon x,e^{q\varepsilon }t_q) \\ 
e^{1\varepsilon }u_{N-1}(e^\varepsilon x,e^{q\varepsilon }t_q) \\ 
\vdots \\ 
e^{N\varepsilon }u_0(e^\varepsilon x,e^{q\varepsilon }t_q) \\ 
e^{(N+1)\varepsilon }u_{-1}(e^\varepsilon x,e^{q\varepsilon }t_q) \\ 
e^{(N+2)\varepsilon }u_{-2}(e^\varepsilon x,e^{q\varepsilon }t_q)
\end{array}
\right) ,  \label{6}
\end{equation}
so the related infinitesimal generator, i.e. conformal symmetry is 
\begin{equation}
\sigma :=\frac d{d\varepsilon }(\varphi _\varepsilon \cdot u)_{\mid
\varepsilon =0}=\left( 
\begin{array}{c}
0 \\ 
u_{N-1} \\ 
2u_{N-2} \\ 
\vdots \\ 
(N+2)u_{-2}
\end{array}
\right) +x\left( 
\begin{array}{c}
u_N \\ 
u_{N-1} \\ 
u_{N-2} \\ 
\vdots \\ 
u_{-2}
\end{array}
\right) _x+qt_q\left( 
\begin{array}{c}
u_N \\ 
u_{N-1} \\ 
u_{N-2} \\ 
\vdots \\ 
u_{-2}
\end{array}
\right) _{t_q}.  \label{7}
\end{equation}
All admissible reductions preserve this scaling property.

Lax representation $(iv)$ for lattice systems is homogenous with respect to the
following scaling 
\begin{equation}
{\cal E}\rightarrow e^\varepsilon {\cal E},\,\,\,u_{N+\alpha
-i}\rightarrow e^{i\varepsilon }u_{N+\alpha -i},\,\,\,\partial
_{t_q}\rightarrow e^{q\varepsilon }\partial _{t_q},\,\,\,L\rightarrow
e^{(N+\alpha )}L.  \label{8}
\end{equation}
The conformal point transformation and conformal symmetry are 
\[
\varphi _\varepsilon \cdot \left( 
\begin{array}{c}
u_{N+\alpha -1} \\ 
\vdots \\ 
u_0 \\ 
\vdots \\ 
u_\alpha
\end{array}
\right) =\left( 
\begin{array}{c}
e^{1\varepsilon }u_{N+\alpha -1}(n,e^{q\varepsilon }t_q) \\ 
\vdots \\ 
e^{(N+\alpha )\varepsilon }u_0(n,e^{q\varepsilon }t_q) \\ 
\vdots \\ 
e^{N\varepsilon }u_\alpha (n,e^{q\varepsilon }t_q)
\end{array}
\right) , 
\]
\begin{equation}  \label{9}
\end{equation}
\[
\sigma =\left( 
\begin{array}{c}
u_{N+\alpha -1} \\ 
2u_{N+\alpha -2} \\ 
\vdots \\ 
(N+\alpha )u_0 \\ 
\vdots \\ 
Nu_\alpha
\end{array}
\right) +qt_q\left( 
\begin{array}{c}
u_{N+\alpha -1} \\ 
u_{N+\alpha -2} \\ 
\vdots \\ 
u_0 \\ 
\vdots \\ 
u_\alpha
\end{array}
\right) _{t_q}. 
\]

\section{Algorythmic construction of recursion operators}

Developing the idea of Sokolov et al. \cite{s} we construct recursion
operators directly from Lax hierarchy 
\begin{equation}
L_{t_q}=[A_q,L].  \label{9a}
\end{equation}
The only necessary information is the explicit form of $L$ operator and the
ansatz $A_{\overline{q}}=A_{q}P+R$ relating $A$ operators with different $q$,
without any specific information of the coefficients of $A.$ Here, $P$ is
some operator that commutes with $L$ of the form $P=P(L)$ and $R$ is the
remainder, hence
\begin{equation}
L_{t_{\overline{q}}}=L_{t_{q}}P(L)+[R,L].
\label{10a}
\end{equation}
\newline
{\em Remark}
\newline In fact, formulas similar to (\ref{10a}) appeared for the first time in 
80{\em th}, derived on more general level in the frame of r-matrix formalism 
(see for example refs. \cite{1v},\cite{2v},\cite{3},\cite{4}). Nevertheless, they contain 
r-matrices and involve more calculations. The advantage of presented method is its 
simplicity, although the roots in r-matrix theory are evident.
\newline

The case $(i),$ i.e. the Gelfand-Dikii case, was considered in details in
ref. \cite{s}, here we concentrate on the cases $(ii)$ and $(iii)$ for field
systems and on the lattice scalar Lax equations $(iv)$.

Of course, in order to apply recursion operators, found by the method
presented, to the construction of hereditary algebra of symmetries, it is
necessary to verify their hereditary property (\ref{3}).

\subsection{Field systems $(ii)$}

$(N+1)$-field Lax operator and the Lax hierarchy are 
\[
L=\partial _x^N+u_{N-1}\partial _x^{N-1}+...+u_0+\partial _x^{-1}u_{-1},
\]
\begin{equation}
L_{t_m}=[A_m,L],\,\,\,\,A_m=(L^{\frac mN})_{\geq 1},\,\,\,m=1,2,...\,\,.
\label{10}
\end{equation}
Let us express the $A_{m+N}$ operator through $A_m,L$ and some remainder $R_m
$ 
\begin{eqnarray}
A_{m+N} &=&(L^{\frac mN}L)_{\geq 1}=(L^{\frac mN}\,_{\geq 1}L+L^{\frac
mN}\,_{<1}L)_{\geq 1}  \nonumber \\
&=&L^{\frac mN}\,_{\geq 1}L-(L^{\frac mN}\,_{\geq 1}L)_0+(L^{\frac
mN}\,_{<1}L)_{\geq 1}  \label{11} \\
&=&A_mL+R_m.  \nonumber
\end{eqnarray}
Analysing the highest and lowest order terms of $R_m$ we conclude that the
remainder is a purely differential operator 
\begin{equation}
R_m=a_m+b_m\partial _x+...+\gamma _m\partial _x^N,  \label{12}
\end{equation}
hence 
\begin{equation}
L_{t_{m+N}}=L_{t_m}L+[R_m,L].  \label{13}
\end{equation}
The right hand side of eq. (\ref{13}) is the pseudodifferential operator: $%
RHS=v_{2N-1}\partial _x^{2N-1}+...+v_0+\partial _x^{-1}v_{-1}+v_{-2}\partial
_x^{-1}u_{-1}.$ From $v_k=0$ for $k=2N-1,...,N,-2,$ we determine $R_m$
coefficients in terms of the coefficients of operators $L$ and $L_{t_m}$.
Comparing the remaining coefficients of both sides of operator equation (\ref
{13}) we get the recurrence formula 
\begin{equation}
\left( 
\begin{array}{c}
u_{N-1} \\ 
\vdots  \\ 
u_{-1}
\end{array}
\right) _{t_{m+N}}=\phi \left( 
\begin{array}{c}
u_{N-1} \\ 
\vdots  \\ 
u_{-1}
\end{array}
\right) _{t_m}.  \label{14}
\end{equation}
There are $N$ commuting chains of vector fields generated by the recursion
operator $\phi $ 
\begin{equation}
K_{r,n}=\phi ^nK_r,\,\,\,\,r=1,...,N,\,\,\,n=0,1,2,...\,\,.  \label{15}
\end{equation}

Let us pass to the admissible constraints. There are few of them \cite{ko},
\cite{b1}. The first constraint is of the form $u_{-1}=0$ and is
non-Hamiltonian. It means that we cannot apply a bi-Hamiltonian method for
construction a recursion operator and so the  method presented becomes even
more important. The second constraint takes the form $u_{-1}=u_0=0$ and
preserves the Hamiltonian structure. The last constraint is the so called
Kupershmidt reduction \cite{k1} 
\begin{equation}
L=(-1)^N\partial _x^{-1}L^{\dagger }\partial _x,\,\,\,q-odd\,\,\,\Rightarrow
\,\,u_{N-1}=0,  \label{16}
\end{equation}
where $(a\partial _x^n)^{\dagger }=(-1)^n\partial _x^na.$ Notice that under
this constraint half of equations from the hierarchy (\ref{9a}) disappear
and only these with odd $q$ remain. Two cases have to be considered. The
first one with an even $N$ preserves the recurrence formula (\ref{13}) as
odd $m$ and even $N$ give odd $m+N$. The second case of odd $N$ is more
complex as $m+N$ is even and so it is excluded from the hierarchy. In this
case we must take 
\begin{eqnarray}
A_{m+2N} &=&(LL^{\frac mN}L)_{\geq 1}=(LL^{\frac mN}\,_{\geq 1}L)_{\geq
1}+(LL^{\frac mN}\,_{<1}L)_{\geq 1}  \nonumber \\
&=&LL^{\frac mN}\,_{\geq 1}L-(LL^{\frac mN}\,_{\geq 1}L)_0-(LL^{\frac
mN}\,_{\geq 1}L)_{-1}+(LL^{\frac mN}\,_{<1}L)_{\geq 1}  \nonumber  \label{17}
\\
&=&LA_mL+R_m,  \label{17a}
\end{eqnarray}
where $R_m=a_m\partial _x^{2N-1}+...+\partial _x^{-1}\gamma _m.$ In
derivation of the formula (\ref{17a}) we applied the identity $\partial
_x^{-1}a_x\partial _x^{-1}=a\partial _x^{-1}-\partial _x^{-1}a$ and the fact
that $u_{N-1}=0.$ Hence, the new recurrence formula is 
\begin{equation}
L_{t_{m+2N}}=LL_{t_m}L+[R_m,L],\,\,\,\,\,\,\,m-odd.  \label{18}
\end{equation}

{\em Example 1. }Consider Lax operator of the Kaup-Broer system 
\[
L=\partial _x+u+\partial _x^{-1}v,\,
\]
with the remainder $R_m=a_m\partial _x+b_m,$ hence 
\[
L_{t_{m+1}}=L_{t_m}L+[R_m,L]
\]
\[
\Updownarrow 
\]
\begin{eqnarray*}
u_{t_{m+1}}+\partial _x^{-1}v_{t_{m+1}} &=&[u_{t_m}-(a_m)_x]\partial
_x+[uu_{t_m}+v_{t_m}+a_mu_x-(b_m)_x] \\
&&+\partial _x^{-1}[uv_{t_m}-(v_{t_m})_x-v(D_x^{-1}v_{t_m})+(a_mv)_x] \\
&&+[u_{t_m}+(D_x^{-1}v_{t_m})+b_m]\partial _x^{-1}v.
\end{eqnarray*}
Finally 
\[
\lbrack u_{t_m}-(a_m)_x]=0\,\,\Rightarrow \,\,\,a_m=D_x^{-1}u_{t_m},
\]
\[
\lbrack u_{t_m}+(D_x^{-1}v_{t_m})+b_m]=0\,\,\Rightarrow
\,\,\,b_m=-u_{t_m}-D_x^{-1}v_{t_m}
\]
and the recursion formula (\ref{14}) takes the form 
\[
\left( 
\begin{array}{c}
u \\ 
v
\end{array}
\right) _{t_{m+1}}=\left( 
\begin{array}{cc}
D_x+D_xuD_x^{-1} & 2 \\ 
v+D_xvD_x^{-1} & -D_x+u
\end{array}
\right) \left( 
\begin{array}{c}
u \\ 
v
\end{array}
\right) _{t_m}.
\]
Imposing the constraint $v=0$ we get the Burgers hierarchy with 
\[
L=\partial _x+u,\,\,\,\,a_m\rightarrow D_x^{-1}u_{t_m},\,\,\,b_m\rightarrow
-u_{t_m}
\]
and the recursion operator of the form 
\[
\phi =D_x+u+u_xD_x^{-1}=D_x+D_xuD_x^{-1}.
\]
The Kupershmidt reduction with odd $N=1$ gives the nonstandard KdV Lax
representation 
\[
L=\partial _x+\partial _x^{-1}v
\]
with Lax recursion 
\[
L_{t_{m+2}}=LL_{t_m}L+[R_m,L],\,\,\,\,\,\,\,\,m-odd,
\]
where 
\[
R_m=a_m\partial _x+b_m+\partial _x^{-1}c_m,
\]
\[
a_m=D_x^{-1}v_{t_m},\,\,\,b_m=0,\,\,\,c_m=-D_xv_{t_m}-vD_x^{-1}v_{t_m}
\]
and related vector field recursion 
\[
v_{t_{m+2}}=(D_x^2+4v+2v_xD_x^{-1})v_{t_m}.
\]

{\em Example 2. }Consider a three field Lax operator 
\[
L=\partial _x^2+u\partial _x+v+\partial _x^{-1}w.
\]
The appropriate remainder takes the form 
\[
R_m=a_m\partial _x^2+b_m\partial _x+c_m,
\]
where 
\[
a_m=\frac 12D_x^{-1}u_{t_m},\,\,\,b_m=\frac 12uD_x^{-1}u_{t_m}-\frac
14D_x^{-1}uu_{t_m}+\frac 12D_x^{-1}v_{t_m},\,\,
\]
\[
c_m=-v_{t_m}-D_x^{-1}w_{t_m}.
\]
So the recurrence formula (\ref{14}) is 
\[
\left( 
\begin{array}{c}
u \\ 
v \\ 
w
\end{array}
\right) _{t_{m+2}}=\left( 
\begin{array}{ccc}
A & \frac 32D_x+\frac 12D_xuD_x^{-1} & 3 \\ 
B & C & 2u \\ 
D & \frac 32w+\frac 12w_xD_x^{-1} & D_x^2-D_xu+v
\end{array}
\right) \left( 
\begin{array}{c}
u \\ 
v \\ 
w
\end{array}
\right) _{t_m},
\]
where 
\begin{eqnarray*}
A &=&\frac 14D_x^2+D_xvD_x^{-1}-\frac 14D_xuD_x^{-1}u, \\
B &=&\frac 34v_x+\frac 23w+(w_x+\frac 12v_{xx}+\frac 12uv_x)D_x^{-1}-\frac
14v_xD_x^{-1}u, \\
C &=&D_x^2+uD_x+v+\frac 12v_xD_x^{-1}, \\
D &=&-\frac 34wD_x+\frac 14uw-\frac 54w_x+\frac
12[(uw)_x-w_{xx}]D_x^{-1}-\frac 14w_xD_x^{-1}u
\end{eqnarray*}
and hence two commuting hierarchies of vector fields are 
\[
K_{1,n}=\phi ^nK_1,\,\,\,\,\,\,K_1=\left( 
\begin{array}{c}
u_x \\ 
v_x \\ 
w_x
\end{array}
\right) ,
\]
\[
K_{2,n}=\phi ^nK_2,\,\,\,\,\,K_2=\left( 
\begin{array}{c}
2v_x \\ 
v_{xx}+2w_x+uv_x \\ 
-w_{xx}+(uw)_x
\end{array}
\right) .
\]
The first reduction $w=0$ leads to $L=\partial _x^2+u\partial _x+v,$ $%
c_m=-v_{t_m}$ ($a_m,\,b_m$ are unchanged), 
\[
\phi =\left( 
\begin{array}{cc}
\frac 14D_x^2+D_xvD_x^{-1}-\frac 14D_xuD_x^{-1}u & \frac 32D_x+\frac
12D_xuD_x^{-1} \\ 
\frac 34v_x+(\frac 12v_{xx}+\frac 12uv_x)D_x^{-1}-\frac 14v_xD_x^{-1}u & 
D_x^2+uD_x+v+\frac 12v_xD_x^{-1}
\end{array}
\right) ,
\]

\[
K_1=\left( 
\begin{array}{c}
u_x \\ 
v_x
\end{array}
\right) ,\,\,\,K_2=\left( 
\begin{array}{c}
2v_x \\ 
v_{xx}+uv_x
\end{array}
\right) . 
\]
The second reduction gives the MKdV hierarchy 
\[
L=\partial _x^2+u\partial _x, 
\]
\[
a_m=\frac 12D_x^{-1}u_{t_m},\,\,\,b_m=\frac 12uD_x^{-1}u_{t_m}-\frac
14D_x^{-1}uu_{t_m},\,\,\,c_m=0, 
\]
\[
\phi =\frac 14D_x^2-\frac 14D_xuD_x^{-1}u,\,\,\,\,K_1=u_x,\,\,\,K_2=0. 
\]
Finally, the Kupershmidt reduction gives another nonstandard KdV case 
\[
L=\partial _x^2+v-\frac 12\partial _x^{-1}v_x, 
\]
where 
\[
R_m=a_m\partial _x+b_m,\,\,\,\,a_m=\frac 12D_x^{-1}v_{t_m},\,\,\,b_m=-\frac
12v_{t_m}, 
\]
and 
\[
\phi =D_x^2+v+\frac 12v_xD_x^{-1}. 
\]
Notice that $L=\phi ^{\dagger }$ and the rescaling $v\rightarrow 4v$ gives
the KdV case from the previous example.

\subsection{Field systems $(iii)$}

$(N+3)$-field Lax operator and the Lax hierarchy are 
\[
L=u_N^N\partial _x^N+u_{N-1}\partial _x^{N-1}+...+u_0+\partial
_x^{-1}u_{-1}+\partial _x^{-2}u_{-2},
\]
\[
L_{t_m}=[A_m,L],\,\,\,\,A_m=(L^{\frac mN})_{\geq 2},\,\,\,m=1,2,...\,\,.
\]
Let us express the $A_{m+N}$ operator through $A_m,L$ and some remainder $R_m
$ 
\begin{eqnarray}
A_{m+N} &=&(L^{\frac mN}L)_{\geq 2}=(L^{\frac mN}\,_{\geq 2}L+L^{\frac
mN}\,_{<2}L)_{\geq 2}  \nonumber \\
&=&L^{\frac mN}\,_{\geq 2}L-(L^{\frac mN}\,_{\geq 2}L)_0-(L^{\frac
mN}\,_{\geq 2}L)_1+(L^{\frac mN}\,_{<2}L)_{\geq 2}  \label{19} \\
&=&A_mL+R_m.  \nonumber
\end{eqnarray}
Analysing the highest and lowest order terms of $R_m$ we conclude that the
remainder is again a purely differential operator 
\begin{equation}
R_m=a_m\partial _x^{N+1}+b_m\partial _x^N+...+\gamma _m,  \label{20}
\end{equation}
hence 
\begin{equation}
L_{t_{m+N}}=L_{t_m}L+[R_m,L].  \label{21}
\end{equation}

There are several admissible constraints. The first group is the following: $%
u_{-2}=0$, $u_{-2}=u_{-1}=0$, $u_{-2}=u_{-1}=u_0=0$, $%
u_{-2}=u_{-1}=u_0=u_1=0.$ The first three are non-Hamiltonian the last one
is Hamiltonian. Another constraint is related with the Kupershmidt reduction 
\begin{equation}
L=(-1)^N\partial _x^{-2}L^{\dagger }\partial _x^2,\,\,  \label{22}
\end{equation}
which survievs only odd terms from the hierarchy. Again an even $N$
preserves the recurence formula (\ref{21}) as odd $m$ and even $N$ give odd $%
m+N$, while for odd $N$ $m+N$ is even and so it is excluded from the
hierarchy. In this case we must take 
\begin{eqnarray}
A_{m+2N} &=&(LL^{\frac mN}L)_{\geq 2}=(LL^{\frac mN}\,_{\geq 2}L)_{\geq
2}+(LL^{\frac mN}\,_{<2}L)_{\geq 2}  \nonumber \\
&=&LL^{\frac mN}\,_{\geq 2}L-(LL^{\frac mN}\,_{\geq 2}L)_1-...-(LL^{\frac
mN}\,_{\geq 2}L)_{-2}+(LL^{\frac mN}\,_{<2}L)_{\geq 2}  \nonumber  \label{23}
\\
&=&LA_mL+R_m,  \label{23a}
\end{eqnarray}
where $R_m=a_m\partial _x^{2N+1}+...+\partial _x^{-2}\gamma _m.$ Hence, the
new recurence formula is 
\begin{equation}
L_{t_{m+2N}}=LL_{t_m}L+[R_m,L],\,\,\,\,\,\,\,\,m-odd.  \label{24}
\end{equation}

{\em Example 3. }Consider $4$-field Lax operator 
\[
L=u\partial _x+v+\partial _x^{-1}w+\partial _x^{-2}z.
\]
The remainder and the recursion operator are 
\[
R_m=a_m\partial _x^2+b_m\partial _x+c_m,
\]
\[
a_m=u^2D_x^{-1}u^{-2}u_{t_m},\,\,%
\,b_m=2D_x^{-3}z_{t_m}+D_x^{-2}w_{t_m}-u_{t_m},\,\,\,
\]
\[
c_m=-D_x^{-2}z_{t_m}-D_x^{-1}w_{t_m}-v_{t_m},
\]
\[
\phi =\left( 
\begin{array}{cccc}
A & u & u_xD_x^{-2}-uD_x^{-1} & 2u_xD_x^{-3}-2uD_x^{-2} \\ 
B & v+uD_x & 2u+v_xD_x^{-2} & 2v_xD_x^{-3}+uD_x^{-1} \\ 
C & w & v-D_xu+w_xD_x^{-2}+wD_x^{-1} & u+2w_xD_x^{-3}+2wD_x^{-2} \\ 
D & z & z_xD_x^{-2}+2zD_x^{-1} & E
\end{array}
\right) ,
\]

\begin{eqnarray*}
A &=&v+uD_x+(u_{xx}+2v_x)u^2D_x^{-1}u^{-2}, \\
B &=&2w+(v_{xx}+2w_x)u^2D_x^{-1}u^{-2}+2wuu_xD_x^{-1}u^{-2}, \\
C &=&3z-3w_x-2wD_x-2wu^{-1}u_x+[2(u^2z)_x-(u^2w)_{xx}]D_x^{-1}u^{-2}, \\
D &=&-3z_x-2zD_x-2zu^{-1}u_x-(u^2z)_{xx}D_x^{-1}u^{-2}, \\
E &=&v-D_xu-wD_x^{-1}+4zD_x^{-2}+2z_xD_x^{-3}.
\end{eqnarray*}
The hierarchy takes the form 
\[
K_{n+1}=\phi ^nK_1,\,\,\,\,\,\,\,\,K_1=\left( 
\begin{array}{c}
u^2u_{xx}+2u^2v_x \\ 
u^2v_{xx}+2u(uw)_x \\ 
-(u^2w)_{xx}+2(u^2z)_x \\ 
-(u^2z)_{xx}
\end{array}
\right) . 
\]
The first constraint $z=0,\,\,L=u\partial _x+v+\partial _x^{-1}w$ gives 
\[
a_m=u^2D_x^{-1}u^{-2}u_{t_m},\,\,\,b_m=D_x^{-2}w_{t_m}-u_{t_m},\,\,%
\,c_m=-D_x^{-1}w_{t_m}-v_{t_m}, 
\]
\[
\phi =\left( 
\begin{array}{ccc}
A & u & u_xD_x^{-2}-uD_x^{-1} \\ 
B & v+uD_x & 2u+v_xD_x^{-2} \\ 
\begin{array}{c}
-3w_x-2wD_x-2wu^{-1}u_x \\ 
-(u^2w)_{xx}D_x^{-1}u^{-2}
\end{array}
& w & v-D_xu+w_xD_x^{-2}+wD_x^{-1}
\end{array}
\right) , 
\]
\[
K_{n+1}=\phi ^nK_1,\,\,\,\,\,\,\,\,K_1=\left( 
\begin{array}{c}
u^2u_{xx}+2u^2v_x \\ 
u^2v_{xx}+2u(uw)_x \\ 
-(u^2w)_{xx}
\end{array}
\right) . 
\]
The second constraint $z=w=0,\,\,L=u\partial _x+v$ gives 
\[
a_m=u^2D_x^{-1}u^{-2}u_{t_m},\,\,\,b_m=-u_{t_m},\,\,\,c_m=-v_{t_m}, 
\]
\[
\phi =\left( 
\begin{array}{cc}
v+uD_x+(u_{xx}+2v_x)u^2D_x^{-1}u^{-2} & u \\ 
v_{xx}u^2D_x^{-1}u^{-2} & v+uD_x
\end{array}
\right) , 
\]
\[
K_{n+1}=\phi ^nK_1,\,\,\,\,\,\,\,\,K_1=\left( 
\begin{array}{c}
u^2u_{xx}+2u^2v_x \\ 
u^2v_{xx}
\end{array}
\right) . 
\]
Finally the third constraint $z=w=v=0,\,\,L=u\partial _x$ gives 
\[
a_m=u^2D_x^{-1}u^{-2}u_{t_m},\,\,\,b_m=-u_{t_m},\,\,\,c_m=0, 
\]
\[
\phi =uD_x+u_{xx}u^2D_x^{-1}u^{-2}, 
\]
\[
K_{n+1}=\phi ^nK_1,\,\,\,\,\,\,\,\,K_1=u^2u_{xx}. 
\]

\subsection{Lattice systems $(iv)$}

$N$-field Lax operator and the Lax hierarchy are 
\[
L={\cal E}^{N+\alpha }+u_{N+\alpha -1}{\cal E}^{N+\alpha
-1}+...+u_\alpha {\cal E}^\alpha ,\,\,\,\alpha =-1,...,-N,\,\,N\geq 2,
\]
\[
L_{t_m}=[A_m,L],\,\,\,\,A_m=(L^{\frac m{N+\alpha }})_{\geq
0},\,\,\,m=1,2,...\,\,.
\]
Let us express the $A_{m+N+\alpha }$ operator through $A_m,L$ and some
remainder $R_m$ 
\begin{eqnarray}
A_{m+N+\alpha } &=&(L^{\frac m{N+\alpha }}L)_{\geq 0}=(L^{\frac m{N+\alpha
}}\,_{\geq 0}L+L^{\frac m{N+\alpha }}\,_{<0}L)_{\geq 0}  \nonumber \\
&=&L^{\frac m{N+\alpha }}\,_{\geq 0}L-(L^{\frac m{N+\alpha }}\,_{\geq
0}L)_{-1}-...-(L^{\frac m{N+\alpha }}\,_{\geq 0}L)_\alpha +(L^{\frac
m{N+\alpha }}\,_{<0}L)_{\geq 0}  \nonumber  \label{25} \\
&=&A_mL+R_m.  \label{25a}
\end{eqnarray}
Analysing the highest and lowest order terms of $R_m$ we conclude that the
remainder is a shift operator of the form 
\begin{equation}
R_m=a_m{\cal E}^{N+\alpha -1}+...+\gamma _m{\cal E}^\alpha ,
\label{26}
\end{equation}
hence 
\begin{equation}
L_{t_{m+N+\alpha }}=L_{t_m}L+[R_m,L].  \label{27}
\end{equation}

There are two possible reductions. The first one takes the form 
\begin{equation}
u_\alpha \neq 0,\,u_{\alpha +1}=...=u_{N+\alpha -1}=0  \label{28}
\end{equation}
\[
L_{t_m}=[L^m\,_{\geq 0},L],\,\,\,m=(N+1)n,\,\,\,n=0,1,2,...\,, 
\]
and includes all Bogoyavlensky lattices, while the second one is 
\begin{equation}
u_0=u_{-1}=...=u_\alpha =0,\,\,\,u_1,...,u_{N+\alpha -1}\neq 0  \label{29}
\end{equation}
\[
L_{t_m}=[A_m,L],\,\,\,\,A_m=(L^{\frac mN})_{\geq 1},\,\,\,m=1,2,...\,\, 
\]
and is a discrete Gelfand-Dikii analog.

We present the calculations on simplest well known examples of Toda and
Voltera lattices.

{\em Example 4. }Consider infinite Toda lattice $N=2,\,\alpha =-1.$ Hence,
we have 
\[
L={\cal E}+p+v{\cal E}^{-1},\,\,\,R_m=a_m+b_m{\cal E}^{-1}
\]
and the Lax hierarchy (\ref{27}) 
\begin{eqnarray*}
p_{t_{m+1}}+v_{t_{m+1}}{\cal E}^{-1} &=&[p_{t_m}+a_m-(Ea_m)]{\cal E}%
+[pp_{t_m}+v_{t_m}+b_m-(Eb_m)] \\
&&+[vp_{t_m}+(E^{-1}p)v_{t_m}+va_m-v(E^{-1}a_m)+b_m(E^{-1}p) \\
&&-pb_m]{\cal E}^{-1}+[(E^{-1}v)v_{t_m}+b_m(E^{-1}v)-v(E^{-1}b_m)]%
{\cal E}^{-2}.
\end{eqnarray*}
Introducing the operator $\Delta :=E-1$ and its inverse $\Delta ^{-1}$ such
that $\Delta ^{-1}f(n)=\sum_{k=-\infty }^{n-1}f(k),\,$we find 
\[
p_{t_m}+a_m-(Ea_m)=0\,\,\Rightarrow \,\,a_m=\Delta ^{-1}p_{t_m},
\]
\[
(E^{-1}v)v_{t_m}+b_m(E^{-1}v)-v(E^{-1}b_m)=0\,\,\,\Rightarrow
\,\,b_m=-v\Delta ^{-1}Ev^{-1}v_{t_m}
\]
and then 
\begin{eqnarray*}
\phi  &=&\left( 
\begin{array}{cc}
(E^{-1}p)+v(E^{-1}\Delta p)\Delta ^{-1}Ev^{-1} & v(1+E^{-1}) \\ 
1+\Delta v\Delta ^{-1}Ev^{-1} & p
\end{array}
\right)  \\
&=&\left( 
\begin{array}{cc}
vE^{-1}\Delta p\Delta ^{-1}Ev^{-1} & v(1+E^{-1}) \\ 
(Ev-vE^{-1})\Delta ^{-1}Ev^{-1} & p
\end{array}
\right) ,
\end{eqnarray*}
\[
K_{n+1}=\phi ^nK_1,\,\,\,\,\,\,\,\,K_1=\left( 
\begin{array}{c}
v(n)[p(n)-p(n-1)] \\ 
v(n+1)-v(n)
\end{array}
\right) .
\]
The constraint $p=0$ leeds to the infinite Volterra system 
\[
L={\cal E}+v{\cal E}^{-1},\,\,\,\,L_{t_{2m}}=[L^{2m}\,_{\geq
0},L],\,\,\,m=1,2,...\,\,.
\]
Now we find 
\begin{eqnarray*}
A_{2m+2} &=&(L^{2m}\,L^2)_{\geq 0}=(L^{2m}\,_{\geq
0}L^2+L^{2m}\,_{<0}L^2)_{\geq 0} \\
&=&L^{2m}\,_{\geq 0}L^2-(L^{2m}\,_{\geq 0}L^2)_{-2}+(L^{2m}\,_{<0}L^2)_{\geq
0} \\
&=&A_{2m}L^2+R_{2m},
\end{eqnarray*}
where $R_{2m}=a_m+b_m{\cal E}^{-2},$ so the recursion chain is 
\[
L_{t_{2m+2}}=L_{t_{2m}}L^2+[R_{2m},L]
\]
and hence 
\[
a_m=\Delta ^{-1}v_{t_{2m}},\,\,\,b_m=-v(E^{-1}v)\Delta
^{-1}Ev^{-1}v_{t_{2m}},
\]
\[
\phi =v+(E^{-1}v)+vE^{-1}+\Delta v(E^{-1}v)\Delta
^{-1}Ev^{-1}=v(1+E^{-1})(EvE-v)\Delta ^{-1}v^{-1},
\]
\[
K_1=v(n)[v(n+1)-v(n-1)].
\]


\begin{thebibliography}{9}
\bibitem{f}  A. S. Fokas and R. L. Anderson: J. Math. Phys. {\bf 23 }(1982)
1066

\bibitem{a}  M. Antonowicz and A. P. Fordy: Comm. Math. Phys. {\bf 124 }%
(1989) 2269

\bibitem{tu} G. Tu: J.Math. Phys. {\bf 30} (1989) 330

\bibitem{g}  M. Gurses and A. Karasu: J. Math. Phys. {\bf 36 }(1995) 3485

\bibitem{b1}  M. B\l aszak: {\em Multi-Hamiltonian Theory of Dynamical Systems, }%
Springer, Berlin 1998

\bibitem{s}  M. Gurses, A. Karasu and V. Sokolov: J. Math. Phys. {\bf 40 }%
(1999) 6473

\bibitem{gd}  I. M. Gelfand and L.A. Dikii: Funct. Anal. Appl. {\bf 10 }%
(1976) 259

\bibitem{k1}  B. A. Kupershmidt: Commu. Math. Phys. {\bf 99 }(1988) 51

\bibitem{b2}  M. B\l aszak and K. Marciniak: J. Math. Phys. {\bf 35 }(1994) 4661

\bibitem{1v}  A. G. Reymann and M. A. Semenov-Tian-Shansky: Funct. Anal. Appl. {\bf 17} (1983) 259

\bibitem{2v}  A. G. Reymann and M. A. Semenov-Tian-Shansky: Publ. RIMS, Kyoto Univ. {\bf 21} (1985) 1237

\bibitem{3}  V. G. Drinfeld: Soviet Math. Dokl. {\bf 27} (1983) 68

\bibitem{4}  W. Oevel: J. Math. Phys. {\bf 30} (1989) 1140

\bibitem{ko}  B. Konopelchenko and W. Oevel: Publ. RIMS, Kyoto Univ. {\bf 29} (1993) 581

\end{thebibliography}
\end{document}